# Ad Lucem: The Photon in the MMHT PDFs

**Ricky Nathvani***
*University College London*
*E-mail:* ricky.nathvani.15@ucl.ac.uk

**Robert Thorne**
*University College London*
*E-mail:* robert.thorne@ucl.ac.uk

**Lucian Harland-Lang**
*University of Oxford*
*E-mail:* lucian.harland-lang@physics.ox.ac.uk

**Alan Martin**
*Durham University*
*E-mail:* a.d.martin@durham.ac.uk

We describe the inclusion of the photon as an additional component of the proton's Parton Distribution Functions (PDFs) in the MMHT framework. The input for the photon is adopted from the recent LUXqed determination. We describe the similarities and differences above the input scale with other photon PDF determinations and the contributions to the MMHT photon from both leading twist and higher twist contributions, and their uncertainties. We study the impact of QED effects on the quark and gluon PDFs and the fit quality, and outline our development of an equivalent set of neutron PDFs.

*Speaker.

## 1. Introduction

Much progress has been made since the first attempts at determining the photon content of the proton's Parton Distribution Functions (PDFs). Their phenomenological importance has also increased in the intervening period as the precision physics program at the Large Hadron Collider (LHC) has been underway. Experimental precision is now becoming sensitive to Quantum Electrodynamic (QED) effects in cross section measurements, such as those for Drell-Yan production [1] or Higgs production with an associated electroweak boson [2], despite an $\mathcal{O}(\alpha)$ factor of suppression. Since PDFs, $xf(x,Q^2)$, are an essential component to the relevant cross section calculations for these processes, appropriate steps must be taken to improve their development on the theoretical front as well.

The earliest such publicly available set was MRSTQED2004 [3] which relied on a phenomenological model of photon radiation from the quarks at the input scale of DGLAP evolution of the partons. Subsequent sets developed similar models [4] or attempted to fit the photon to data [5], such as Drell-Yan production at the LHC, usually at the expense of large errors of $\mathcal{O}(100\%)$.

Recently, the determination of the photon from structure functions has been developed by Harland-Lang et al [6] and LUXqed [7] on a more rigorous quantitative basis. This has arisen from the understanding of the correspondence of the photon in Deep Inelastic Scattering (DIS) processes involving proton structure functions and the photon PDF as an interacting parton in proton-proton collisions at the LHC [8].

As discussed in Section 2, we now adopt an approach based on that of LUXqed, and also take account of developments regarding the incorporation of QED in the DGLAP equation at $\mathcal{O}(\alpha)$, $\mathcal{O}(\alpha\alpha_S)$ [9] and $\mathcal{O}(\alpha^2)$ [10] for the proton. Furthermore, we adopt this formalism to produce an equivalent set for the neutron.

## 2. QED in the MMHT Framework

Our expression for the input distribution of the photon, $x\gamma(x,Q_0^2)$, is adapted directly from that of LUXqed [7]. The equation expresses the photon content in terms of $F_2$, $F_L$, allowing the input to be derived from structure function data (for $Q^2 < Q_0^2$) obtained in DIS experiments:

$$x\gamma(x,Q_0^2) = \frac{1}{2\pi\alpha(Q_0^2)}\int_x^1 \frac{dz}{z}\Bigg\{\int_{\frac{x^2m_p^2}{1-z}}^{Q_0^2}\frac{dQ^2}{Q^2}\alpha^2(Q^2)\left[\left(zp_{\gamma,q}(z)+\frac{2x^2m_p^2}{Q^2}\right)F_2(x/z,Q^2)-z^2F_L(x/z,Q^2)\right] \\ -\alpha^2(Q_0^2)\left(z^2F_2(x/z,Q_0^2)-\ln(1-z)(z^2-2z+2)F_2(x/z,Q^2)\right)\Bigg\} \tag{2.1}$$

Likewise, we use data from the CLAS [11], Christy-Bosted [12] and HERMES [13] fits to the inelastic structure function data, $F_2^{inel}$, from HERA, and the A1 collaboration fit [14] for the elastic scattering data, $F_2^{el}$. Our expression differs from that of LUXqed in that the upper limit of the integral in $Q^2$ is taken as $Q_0^2 = 1$ GeV$^2$, the starting scale in MMHT, in order to include the photon from input scale in the simultaneous DGLAP evolution of all the partons. Furthermore, the final term, $\sim \ln(1-z)$, is included to account for the difference in form, since our expression takes the

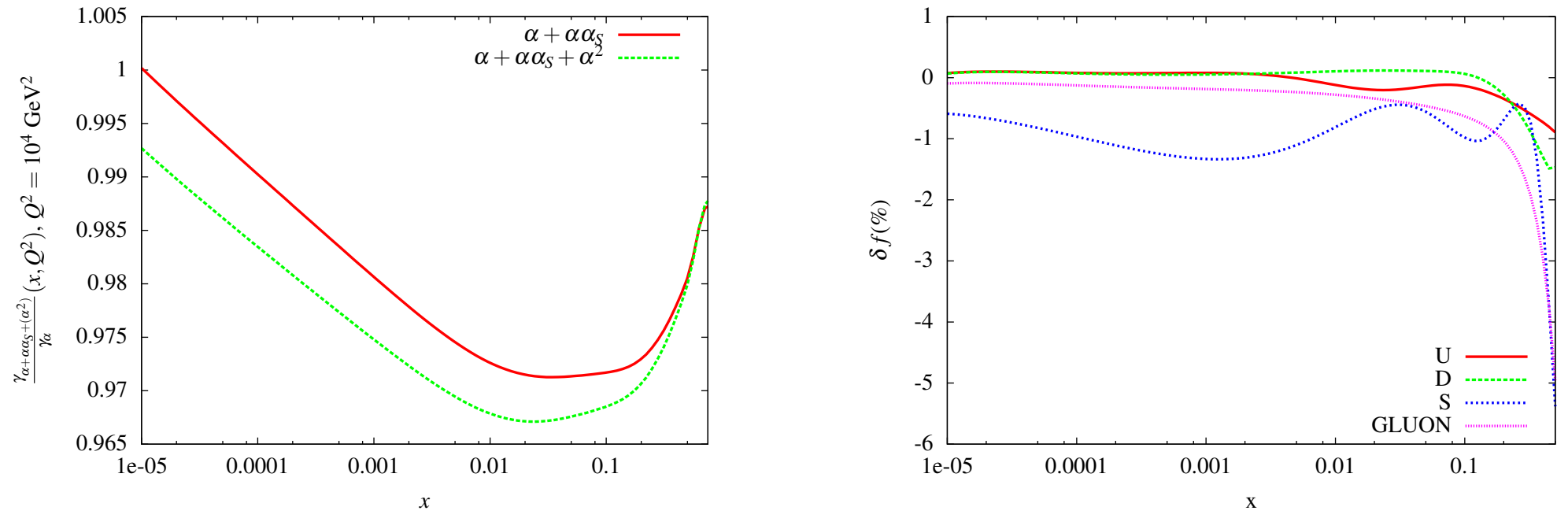

(a) Ratio of the Photon PDF with and without $\mathcal{O}(\alpha\alpha_S)$,$\mathcal{O}(\alpha^2)$ corrections, $Q^2 = 10^4$ GeV$^2$.

(b) QED corrections to $xf(x,Q^2)$, $Q^2 = 10^4$ GeV$^2$, after refitting all partons.

Figure 1

starting scale as the upper limit, in contrast to $\frac{Q_0^2}{1-z}$ as found in LUXqed. Due to the lower scale from which our partons are evolved from compared to LUXqed, certain factors such as the proton mass term $\mathcal{O}(\frac{m_p^2}{Q^2})$, and higher twist terms (see Section 5) are more relevant to our expression, particularly at high $x$. The contribution to the photon from $F_2^{el}$, also known as the coherent contribution, is added in above the input scale, outside of DGLAP evolution, which introduces a negligible amount of momentum violation of $\mathcal{O}(10^{-4})$.

The DGLAP evolution of all the partons is performed with QED splitting kernels $p_{i,j}^{QED}$ of $\mathcal{O}(\alpha)$, $\mathcal{O}(\alpha\alpha_S)$ and $\mathcal{O}(\alpha^2)$ as developed in [9][10]. After performing a refit of the PDF parameters with QED DGLAP evolution, these are seen to have a noticeable impact to the structure of the photon during evolution (fig. 1a), particularly at high-$x$ where the effects of higher order splitting terms are of $\mathcal{O}(1-3\%)$. We also include target mass corrections for the proton alongside the $p_{\gamma,\{q,g\}}^{QED}$, in the same manner as the first term convoluted with $F_2$ in (2.1). Furthermore, since (2.1) holds above the input scale, the lower limit of the integral in $Q^2$ introduces a kinematic cut on all contributions from the partons and the coherent contributions to the photon PDF at high $x$.

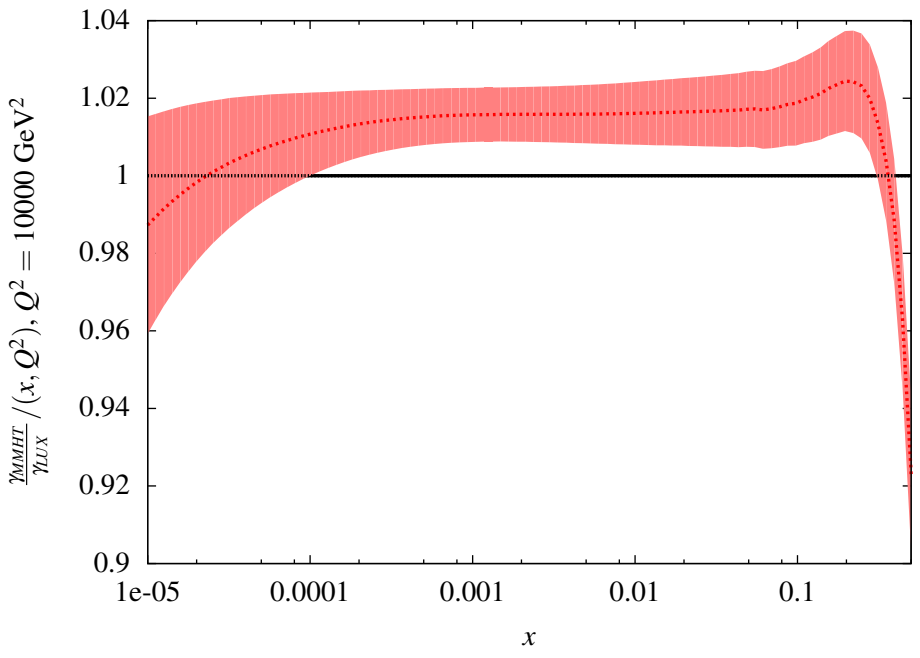

Figure 2: Ratio of photon PDF in MMHT and LUXqed, $Q^2 = 10^4$ GeV$^2$.

As seen in fig. 1b, due to an $\mathcal{O}(\alpha)$ suppression, the effect of QED splitting kernels $p_{i,j}^{QED}$ on the quarks is moderate after refitting the PDFs. The gluon is also affected, since splitting kernels of $\mathcal{O}(\alpha\alpha_S)$ couple it through QED effects. The strange and anti-strange are especially sensitive, since they are less well constrained, particularly at high $x$.

In order to model higher twist effects a model of power corrections to the DIS structure functions is adopted. We take the form proposed by [15] for infrared renormalon contributions that characterise low $Q^2$ divergences, which enhances the photon PDF at high-$x$ (see Section 5). Changes arising from different orders of evolution in QCD have a modest effect on the photon, leading to changes of $\mathcal{O}(1-2\%)$.

While older photon PDFs had larger uncertainties and variance between sets, particularly at high-$x$, those of LUXqed, the more recent NNPDF set [16], which also adopts the LUX input formulation, and our set are now in $\mathcal{O}(1\%)$ agreement in the phenomenologically significant range, as seen in fig. 2.

Below, we provide a table of changes in $\chi^2$ after incorporating QED effects, both with the parton parameters as fit in older sets and after re-fitting the partons entirely with QED in DGLAP evolution. The increase in $\chi^2$ after re-fitting with QED inclusions is found to be primarily to due tension with the BCDMS $F_2$ data.

| Change in $\chi^2$ due to QED evolution compared to MMHT14+HERA I+II | | | |
|---|---|---|---|
| NLO before fit | NLO after fit | NNLO before fit | NNLO after fit |
| +28 | +17 | +29 | +13 |

## 3. Neutron

Efforts have been successfully undertaken by the MMHT group to consistently include the aforementioned effects in the development of a set of neutron PDFs, including a corresponding photon PDF. For the neutron photon PDF, the principal approximation is to assume that isospin symmetry, $(u+\bar{u})_n = (d+\bar{d})_p, (d+\bar{d})_n = (u+\bar{u})_p$ holds for the quarks to a high degree of accuracy in spite of isospin-violating effects introduced by QED to their evolution in DGLAP (an assumption tested below).

At input, the neutron photon expression is identical to (2.1), with the modification that all contributions from $F_2^{el}$ are calculated from fits to neutron structure functions [14] and the target mass term modified to use the neutron mass. However, since the neutron has no net charge, the data is consistent with this contribution being $\simeq 0$ to a good approximation. The inelastic term is then approximated by pre-multiplying the proton contribution by the ratio of charge weighted partons in the neutron to those of the proton.

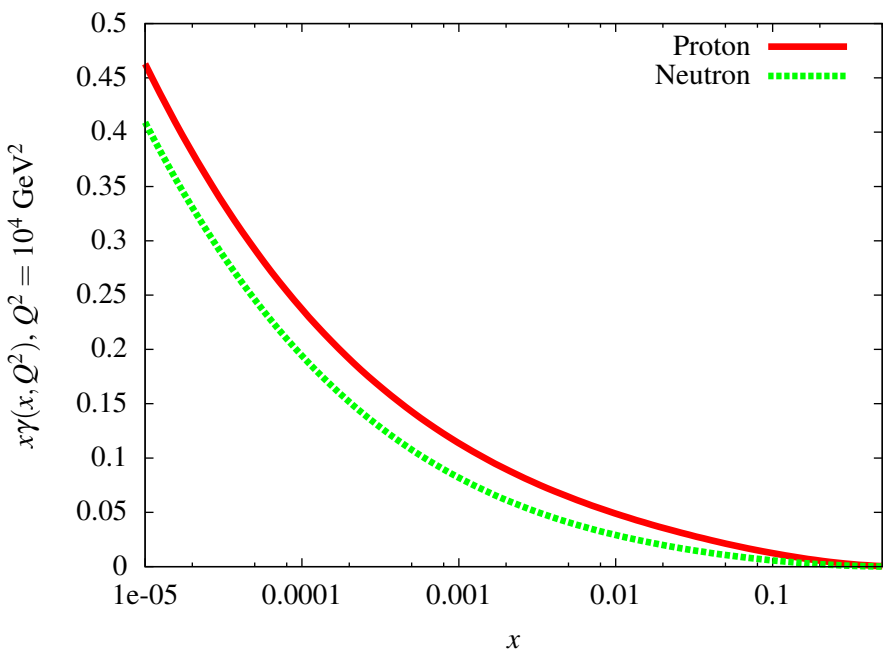

Figure 3: Comparison of the photon PDF in the proton and neutron, $Q^2 = 10^4$ GeV$^2$.

The subsequent contributions to the neutron photon from evolution are estimated in the following way. The photon receives contributions from the quarks in DGLAP as follows:

$$\gamma(x,\mu^2)_{q,p} = \int_{Q_0^2}^{\mu^2} \frac{\alpha(Q^2)}{2\pi} \frac{dQ^2}{Q^2} \int_x^1 \frac{dz}{z} \left( e_q^2 p_{\gamma,q}^{QED}(z) q(\frac{x}{z}, Q^2) \right). \tag{3.1}$$

If one uses the approximation of isospin symmetry, then the neutron photon can be estimated by charge re-weighting the contributions from the $d+\bar{d}$ and $u+\bar{u}$ distributions, assuming that sea quarks remain approximately invariant between the proton and neutron:

$$\gamma(x,\mu^2)_n = \frac{e_d^2}{e_u^2} \gamma_{u,p}(x,\mu^2) + \frac{e_u^2}{e_d^2} \gamma_{d,p}(x,\mu^2) + \gamma_{\{s,c,b,g\},p}(x,\mu^2). \tag{3.2}$$

Although substantially smaller at $Q^2 = 1$ GeV$^2$, at the electroweak scale, the neutron photon PDF is comparable in magnitude to that of the proton, particularly at small $x$ as seen in fig. 3.

For the valence distributions in the neutron, we assume that the isospin-violation is proportional to the contributions to the valence distributions in the proton that arise from the dominant QED splitting kernels in DGLAP:

$$\Delta q(x,\mu^2)^{QED} = \int_{Q_0^2}^{\mu^2} \frac{\alpha(Q^2)}{2\pi}\frac{dQ^2}{Q^2}\int_x^1 \frac{dz}{z}\Big(e_q^2 p_{q,q}^{QED}(z)q(\frac{x}{z},Q^2)\Big). \tag{3.3}$$

Then, the isospin violating terms in the neutron $\Delta d_{V,n}(x,\mu^2) = d_{V,n}(x,\mu^2) - u_{V,p}(x,\mu^2)$, $\Delta u_{V,n}(x,\mu^2) = u_{V,n}(x,\mu^2) - d_{V,p}(x,\mu^2)$, are taken as:

$$\Delta d_{V,n} = \varepsilon(1-\frac{e_d^2}{e_u^2})\Delta u_{V,p}^{QED}, \qquad \Delta u_{V,n} = \varepsilon(1-\frac{e_u^2}{e_d^2})\Delta d_{V,p}^{QED}, \tag{3.4}$$

where $\varepsilon$ is fixed to conserve momentum conversation at input in the neutron. The inclusion of such terms introduces isospin violating effects of $\mathcal{O}(5\%)$, which could have implications for the determination of nuclear PDFs.

## 4. Uncertainties

Because of our adoption of the input expression for the photon from LUXqed, our uncertainties bear strong resemblance to their photon with some exceptions. Similar to LUX, our treatment of the uncertainties for $F_2$ and $F_L$ are taken from the uncertainty bands provided by the fits from A1 and CLAS, as well as comparison to an alternative fit by Christy and Bosted [12] as an estimation of uncertainty of $F_2^{inel}$ in the continuum region.

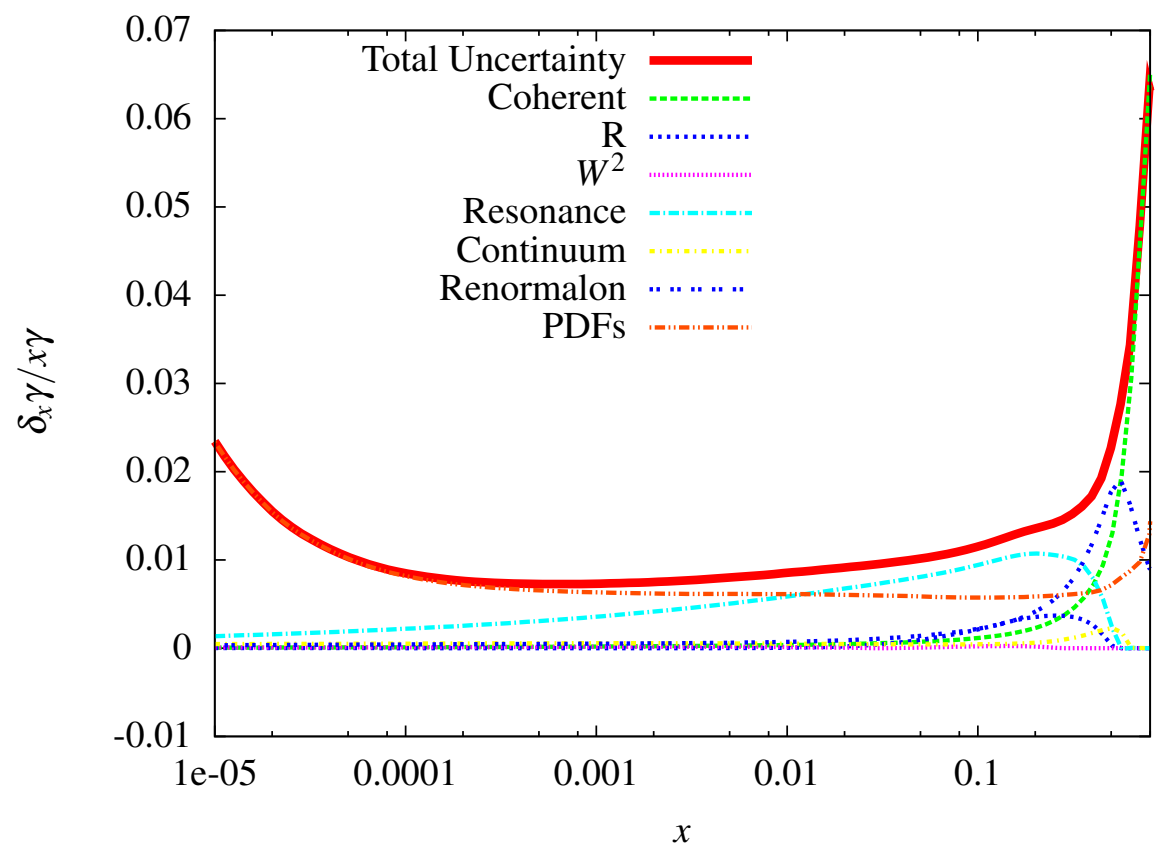

Figure 4: Photon PDF Uncertainty contributions (added in quadrature to give the total uncertainty), $Q^2 = 10^4$ GeV$^2$.

One source of uncertainty for the Inelastic form factor contribution is from the CLAS fit [11] to the resonance $F_2$ (for $W^2 > W_{cut} = 3.5$ GeV$^2$) and the HERMES fit [13] for the continuum $F_2$ data (for $W^2 < W_{cut}$). Since no smoothing is performed between these two regions, our sensitivity to $W_{cut}$ is estimated by varying $3 < W_{cut}^2 < 4$ and is considered as an independent source of uncertainty. In practice, this is found to be negligible. We take the uncertainty bands provided by [13] for the HERMES fit as an independent source of uncertainty.

Above the input $Q_0^2 = 1$ GeV$^2$, the $F_2$ contributions to the photon are handled by the quark to photon splittings in DGLAP. The standard eigenvector uncertainties on the PDFs are then taken as an independent source of uncertainty in the photon, while the uncertainty from the coherent contribution is taken independently from the A1 fit. This does not, however, account for non-perturbative power corrections, closely related to the infrared divergences in perturbation series for field theories such as QCD. One characterisation of these divergences is given by the so called renormalon. The form of the infrared divergence for $F_2$ can be calculated and accounted for in the quark distribution as passed in DGLAP for the photon evolution. The modification to the quark distribution, as calculated by [15] at $O(\frac{1}{Q^2})$ is then as follows:

$$q(x,Q^2) \to q(x,Q^2)\Big(1+A_2'\int_x^1 \frac{dz}{z}C_2(z)q(\frac{x}{z},Q^2)\Big), \tag{4.1}$$

where $C_2$ is defined in [15]. The parameter $A_2'$ is not well determined, being fit loosely to structure function data, and our central value is taken as $A_2' = 0.3^{+0.1}_{-0.1}$, from our own global best fit estimation to structure function data. Overall, the central values and uncertainties of our photon PDF, after simultaneous DGLAP evolution with all partons and accounting for lower $Q^2$ effects, bears a strong resemblance to that of LUXqed, with total uncertainties of $\mathcal{O}(1-2\%)$.